\documentclass[superscriptaddress,reprint,prl,amsmath,amssymb,aps]{revtex4-1}

\usepackage{graphicx}
\usepackage{dcolumn}
\usepackage{bm}
\usepackage{longtable}
\usepackage[T1]{fontenc} 
\usepackage{lipsum} 
\usepackage{amsthm}
\usepackage{amsfonts}
\usepackage{mathtools}
\usepackage{booktabs}
\usepackage{amsmath}
\usepackage{blindtext}
\usepackage{setspace}
\usepackage{makeidx}
\usepackage{rotating}
\usepackage{tikz}
\usepackage{gensymb}
\usepackage{float}

\begin{document}

\title{A new class of half-metallic ferromagnets from first principles}

\author{Sin\'{e}ad M. Griffin}
\affiliation{Molecular Foundry, Lawrence Berkeley National Laboratory, Berkeley, CA 94720, USA}
\affiliation{Department of Physics, University of California Berkeley, Berkeley, CA 94720, USA}
\email{sgriffin@lbl.gov}
\homepage{http://sites.google.com/sineadv0}

\author{Jeffrey B. Neaton}
\affiliation{Molecular Foundry, Lawrence Berkeley National Laboratory, Berkeley, CA 94720, USA}
\affiliation{Department of Physics, University of California Berkeley, Berkeley, CA 94720, USA}
\affiliation{Kavli Energy NanoScience Institute at Berkeley, Berkeley, CA 94720, USA}
\email{jbneaton@lbl.gov}

\date{\today}

\begin{abstract}

Half-metallic ferromagnetism (HMFM) occurs rarely in materials and yet offers great potential for spintronic devices. Recent experiments suggest a class of compounds with the `ThCr$_{2}$Si$_{2}$' (122) structure -- isostructural and containing elements common with Fe pnictide-based superconductors -- can exhibit HMFM. Here we use \textit{ab initio} density-functional theory calculations to understand the onset of half-metallicity in this family of materials and explain the appearance of ferromagnetism at a quantum critical point. We also predict new candidate materials with HMFM and high Curie temperatures through A-site alloying.

\end{abstract}

\maketitle

For the past two decades, the field of spintronics has aimed to augment standard charge-based technologies by harnessing phenomena associated with the spin degree of freedom\cite{Ohno:1998, Bader/Parkin:2010}. The most promising and sought-after spintronic property is half-metallic ferromagnetism owing to its strong potential for applications such as spin-filters\cite{Park_et_al:1998, DeGroot_et_al:1983}. The promise of materials with HMFM is associated with their unusual electronic structure -- they are metallic in one spin channel but insulating in the other -- resulting in total spin polarization at the Fermi level.

Although HMFM has been proposed in oxides\cite{Coey/Venkatesan:2002}, sulfides\cite{Devey_et_al:2009}, Heuslers\cite{DeGroot_et_al:1983} and graphene\cite{Son_et_al:2006}, it is extremely rare. (Ga, Mn)As remains the most-studied system with HMFM, given its potential ease of incorporation into existing semiconductor technologies and the significant capabilities for epitaxy (and therefore tunability) of III-V semiconductors\cite{Dietl_et_al:2000,Shioda_et_al:1998}.  Both spin and charge are introduced into GaAs by substituting trivalent Ga$^{3+}$ with divalent magnetic ions, Mn$^{2+}$, providing ferromagnetically-coupled spin carriers. However its use in devices has been hampered by the low solubility of Mn in GaAs, limiting its transport and magnetic properties, and resulting in Curie temperatures below room temperature. 

Recently, evidence of half-metallic ferromagnetism was found in Ba$_{0.4}$Rb$_{0.6}$Mn$_{2}$As$_{2}$ with a Curie temperature, $T_{C}$, of 103 K\cite{Pandey/Johnston:2015}, Ba$_{0.6}$K$_{0.4}$Mn$_{2}$As$_{2}$ with a $T_{C}$ of 100 K\cite{Pandey_et_al:2013}, and (Ba$_{1-x}$K$_{x}$)(Zn$_{1-y}$Mn$_{y}$)$_{2}$As$_{2}$ with a $T_{C}$ of 180 K\cite{Zhao_et_al:2013}. These form in a tetragonal `ThCr$_{2}$Si$_{2}$' (122) structure, which is isostructural with the high-temperature phase of the (122) Fe-pnictide-based superconductors\cite{Kamihara_et_al:2008, Paglione/Greene:2010}. Varying the transition-metal ion in these 122 pnictide structures leads to a highly-divergent set of phenomena: high-T$_{C}$ superconductivity in Fe-based materials\cite{Rotter_et_al:2008}, antiferromagnetic semiconducting behavior in Mn-based compounds\cite{Singh_et_al:2009, An_et_al:2009}, and non-Fermi-liquid behaviour in YbRh$_{2}$Sr$_{2}$\cite{Custers_et_al:2003}. The potential coexistence of half-metallic ferromagnetism and superconductivity may enhance the existing properties of both phenomena, and result in new states of matter\cite{Linder/Robinson:2015}. Such coexistence in (Fe, Ga)As-based materials has been considered in the `ThCr$_{2}$Si$_{2}$' structure, where they were found to be mutually exclusive due to the nature of the Fe-As bonds\cite{Griffin/Spaldin:2012,Prinz_et_al:1982}. The Fe ions favor a striped antiferromagnetic order in these materials; to achieve HMFM a ferromagnetic ground state is required.

\begin{figure}
 \centering
 \includegraphics[width=0.9\linewidth]{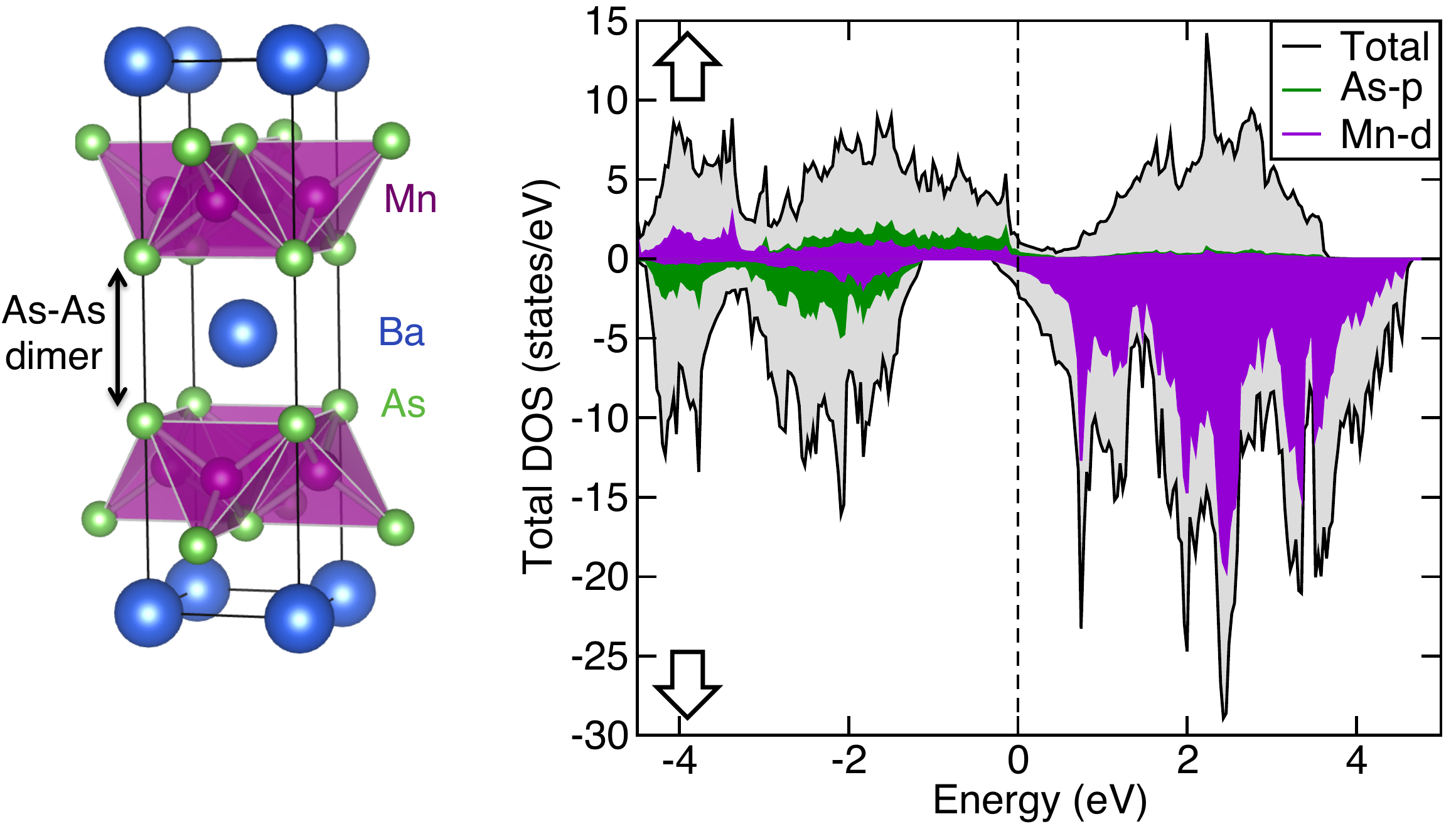}
 \caption{The I4/mmm crystal structure with intralayer As-As dimer. The calculated spin-polarized density of states for stoichiometric BaMn$_{2}$As$_{2}$ with ferromagnetic ordering. The Mn-3d and As-4p orbitals are projected and depicted in purple and green respectively. The spin up channel is plotted as positive, down is negative. The Fermi energy is set to 0 eV. }
 \label{ba_stoich}
\end{figure}

BaMn$_{2}$As$_{2}$ adopts an I4/mmm structure where Mn and As form edge-sharing polyhedra and take up a 2D layered structure separated by layers of large Ba ions. Potential half-metallicity in the bulk I4/mmm structure offers several advantages to Mn-doped GaAs: the synthesis and solubility issues affecting Mn substitution are avoided and there is a range of variants accessible via chemical substitution to explore . The FM ordering required for half-metallicity has been observed in isostructural compound SrCo$_{2}$(Ge$_{1-x}$P$_{x}$)$_{2}$\cite{Jia_et_al:2011}, which exhibits unexpected FM order for $0.35\leq x\leq 0.7$ from low-temperature susceptibility measurements. The saturated moments are low but finite, varying between 0.04 $\mu_{B}$ and 0.2 $\mu_{B}$ per formula unit. The breaking of the intralayer Ge-Ge dimer (analogous to our As-As dimer) with $x$ is suggested to cause a ferromagnetic quantum critical point (QCP). When the dimer is fully broken, the FM state disappears due to a charge redistribution in the Co$_{2}$P$_{2}$ layers, resulting in the new magnetic order. In fact, stoichiometric BaMn$_{2}$As$_{2}$ with a hypothetical FM ordering is close to half-metallic: a shift of the Fermi level by hole doping would potentially result in a HMFM (Fig.\ref{ba_stoich}). This could be achieved by substituting monovalent \textit{A}$^{1+}$ ions on the Ba$^{2+}$ sites to provide hole doping without disrupting the Mn-As layer.

In this letter, we first examine the electronic and magnetic properties of (Ba,Rb)Mn$_2$As$_2$ alloys and explain the observed half-metallicity in the system; and then calculate the structural, electronic and magnetic properties of 15 additional compounds of $A_{x}$$B_{1-x}$Mn$_2$As$_2$ and $A_{x}$$B_{1-x}$Mn$_2$Sb$_2$ ($A$ = group I, $B$ = group II) alloys, prediciting several alloys as potential new HMFMs.

Density functional calculations are performed with the Vienna ab initio Simulation Package (VASP)\cite{VASP1, VASP2} code. We use the PBE functional\cite{PBE1}, with Hubbard U corrections for the Mn-d states. In all DFT+U\cite{Dudarev_et_al:1998} calculations, we use a $U_{eff}$ of 4 eV. Calculations of random alloys that explicitly treat disorder over broad composition ranges remain challenging for DFT-based approaches, as the large unit cells required can become computationally intractable at experimentally-relevant alloy compositions. Here we adopt the virtual crystal approximation (VCA)\cite{Bellaiche/Vanderbilt:2000,Ramer/Rappe:2000}. As the doping is isovalent and does not change the states present at the Fermi energy, the VCA is a good approximation for our alloys. Further calculation details are given in the Supplementary Information (SI) - (I, II). 

All calculations -- both for the stoichiometric and alloyed compounds -- are initialized in the I4/mmm structure and a relaxation of both the lattice and internal coordinates is performed. The relative energy to another possible crystal structure, with P$\bar{3}$m1 symmetry, is also calculated after a full relaxation for alloys of 0\%, 50\% and 100\% composition on the \textit{A} site for all \textit{A}-site combinations considered in this work to determine the stable structure. This P$\bar{3}$m1 structure is the ground-state for CaMn$_{2}$As$_{2}$ and SrMn$_{2}$As$_{2}$. The results for all compounds considered are given in the SI (IV). We begin with a discussion of stoichiometric BaMn$_2$As$_2$ which has a ground-state G-type (checkerboard) antiferromagnetic (AFM) order with a N\'{e}el temperature of 625K\cite{Singh_et_al:2009}. Our DFT calculations confirm G-type AFM order, which we calculate to be 35 meV per formula unit lower in energy than FM. The calculated moment of the Mn ions in the G-type ordering is  $m_{Mn}=4.41 \mu_{B}$, slightly overestimating the experimental value of $m_{Mn}=3.88 \mu_{B}$\cite{Singh_et_al:2009}. A modest reduction in our imposed U by 0.5 eV reproduces this lower experimental moment.

Interestingly, imposing FM ordering on BaMn$_2$As$_2$ causes an exchange splitting between the majority (spin-up) states and the minority (spin-down) states, shown in the calculated density of states (DOS) in Fig.\ref{ba_stoich}(b). Though there are states in both spin channels at the Fermi level, hole doping could shift the Fermi level into a regime where the compound exhibits half-metallicity. The most straightforward route to hole doping is to replace Ba$^{2+}$ with monovalent alkalis (Na, K, Rb, Cs). Such alloying provides the needed holes to shift the Fermi level but avoids disrupting the Mn-As layer which dominates the physics at the Fermi level. Once doping into the half-metallic regime has been achieved, it remains to obtain a FM ground state.

\begin{figure}
 \centering
 \includegraphics[width=0.6\linewidth]{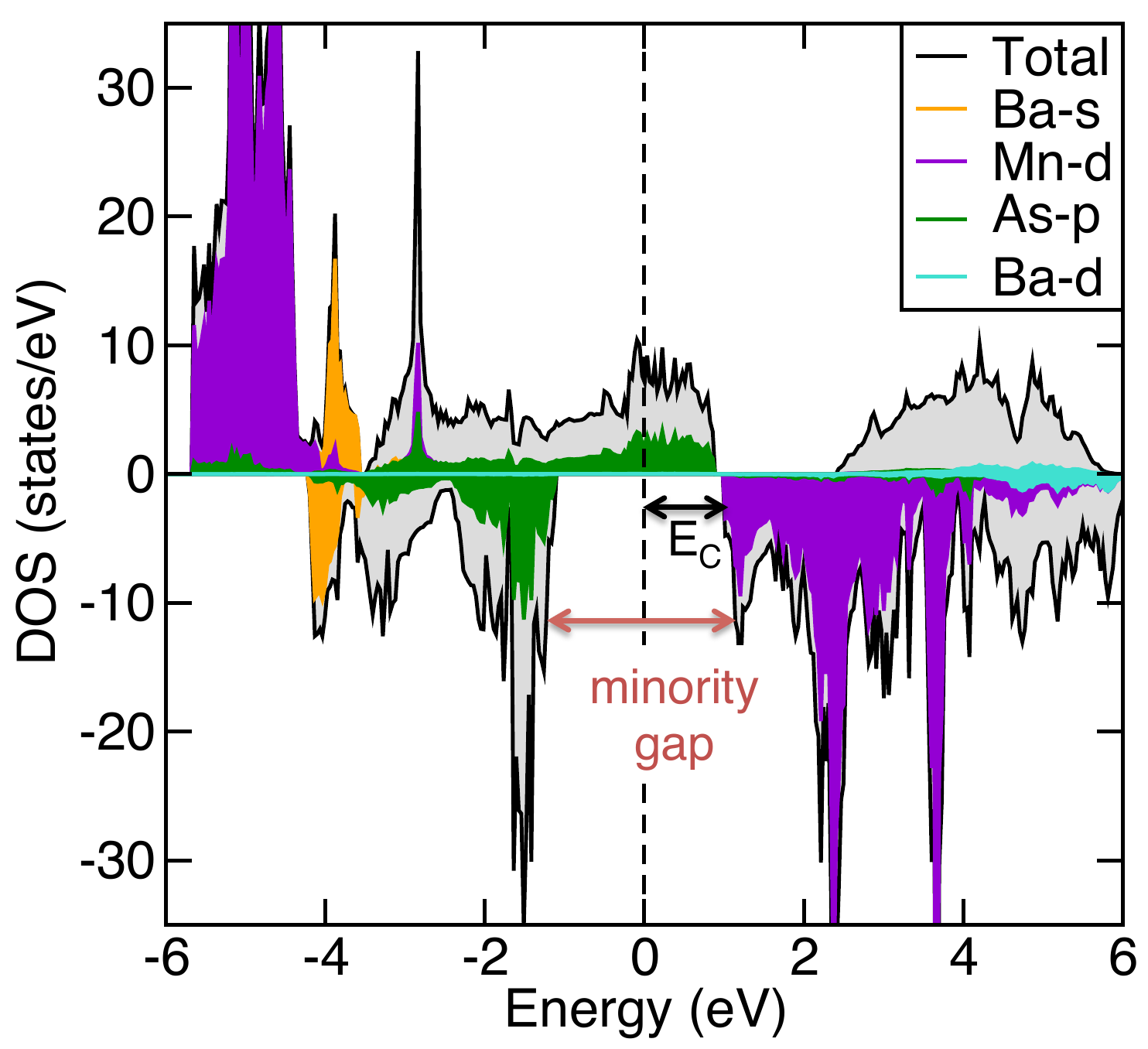}
 \caption{Calculated spin-polarized density of states for Ba$_{0.5}$Rb$_{0.5}$Mn$_{2}$As$_{2}$ with ferromagnetic order. The orbitally-projected states for Ba-s, Ba-d, Mn-d and As-p are shown in orange, blue, purple and green respectively and the Fermi level is set to 0 eV. The spin up channel is plotted as positive, down is negative.}
 \label{5050dos}
\end{figure}

Following recent experiments\cite{Pandey/Johnston:2015}, we initially hole-dope BaMn$_{2}$As$_{2}$ by alloying on the A-site with Rb. The calculated electronic structure for 50\% Rb alloying in Fig.\ref{5050dos} has 100\% spin polarization with the minority gap of 2 eV with near gap states of As-p and Mn-d character. In particular, the conduction band edge is As-p character. We find that for the range of alloys in which FM is preferred, we obtain a half-metallic electronic structure. Additionally, calculated values of $E_{C}$, the difference between the Fermi energy and the minority conduction band minimum, vary from 0.68 eV to 0.99 eV. At room temperature, the value $E_{C} >> kT$ in all cases, making spin contamination\cite{Coey:2010} highly unlikely.

\begin{figure}
 \centering
 \includegraphics[width=0.9\linewidth]{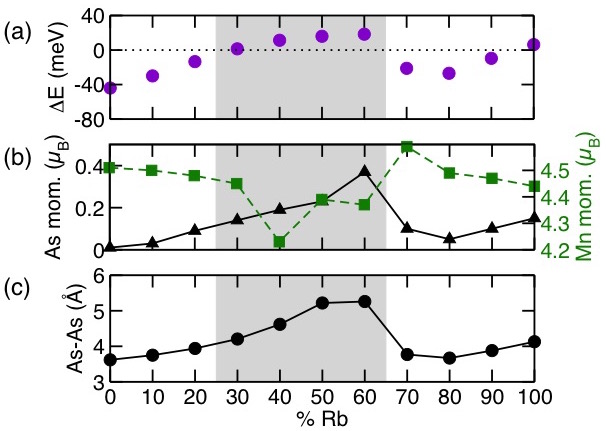}
 \caption{Calculations for FM-ordered Ba$_{1-x}$Rb$_{x}$Mn$_{2}$As$_{2}$ as a function of Rb alloying. (a) shows the calculated energy difference, $\Delta E = E_{AFM}-E_{FM}$ between AFM and FM ordering per formula unit. (b) gives the calculated magnetic moment on As-p (left) and Mn-d (right) orbitals. Finally, (c) shows the calculated intralayer As-As dimer distance, shown in Fig. \ref{qcp}. The shaded region depicts the alloy range when FM is calculated to be lower in energy than AFM order.}
 \label{rbba_only}
\end{figure}

A FM ground state is necessary for a half-metallic electronic structure. We perform VCA calculations of Ba$_{1-x}$Rb$_{x}$Mn$_{2}$As$_{2}$ varying $x$ in steps of 0.1 from 0 to 1. We calculate the energy difference between G-type AFM (the stoichiometric ground state) and FM ordering, as shown in Fig. \ref{rbba_only}(a). For Rb concentrations between 30\% and 60\%, we compute a FM ground state, with a maximum energy difference of 18 meV per formula unit. In fact, $\Delta$E is proportional to the magnitude of exchange coupling and can thus be correlated with $T_{C}$. We see with increased alloying $\Delta$E increases, suggesting that higher alloy percentages will raise $T_{C}$. The experimental value of  $T_{C}$= 103 K for Ba$_{0.4}$Rb$_{0.6}$Mn$_{2}$As$_{2}$\cite{Pandey/Johnston:2015} could thus potentially be increased with greater Rb concentration.

Fig. \ref{rbba_only}(b) shows the calculated magnetic moments due to the Mn-3d and As-4p orbitals for the FM case. The As-p states acquire a nonzero moment which increases from 0  to 0.4 $\mu_{B}$ with greater Rb concentration, which is partially compensated by a slight reduction in the Mn-d moment in this range from 4.5 to 4.2 $\mu_{B}$. In fact, from our calculated density of states of the 50\% Rb alloy in Fig. \ref{5050dos}, the FM charge carriers have As-p character, and so the exchange interactions between itinerant As-p states will be responsible for the observed T$_{C}$ in these alloys.

Our calculations indicate that the As-p moments are most sensitive to the As-As intralayer dimer length, which varies with Rb concentration. Notably, this sensitivity is analogous to the relationship between a similar Ge-Ge dimer and a ferromagnetic QCP in SrCo$_{2}$(Ge$_{1-x}$P$_{x}$)$_{2}$ by Jia et al.\cite{Jia_et_al:2011}. In the present case, the As-p orbitals acquire a non-negligible magnetic moment at a Rb concentration of 20\%, corresponding to an As-As dimer length of 3.94 \AA. As the Rb concentration increases, the As-As distance grows to a maximum of 5.26 \AA, with the As-p moments reaching0.37 $\mu_{B}$. Between 60\% and 70\% Rb alloying, the As-As bond length collapses back to 3.77 \AA, close to its initial value, which is accompanied by a steep drop in the As-p moments to 0.1 $\mu_{B}$. 

\begin{figure}
 \centering
 \includegraphics[width=0.7\linewidth]{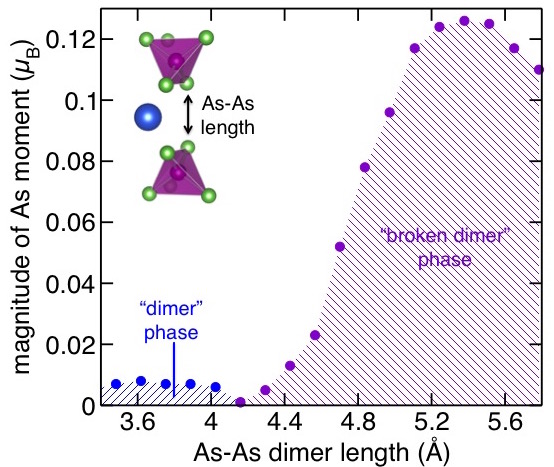}
 \caption{Calculated As-p magnetic moment with changing the depicted As-As dimer length in stoichiometric BaMn$_{2}$As$_{2}$.}
 \label{qcp}
\end{figure}

We illustrate the role of the hole doping on the As-As dimer length by performing calculations in which we systematically change the As-As dimer length in stoichiometric BaMn$_{2}$As$_{2}$ (Fig.\ref{qcp}). The cell volume and shape is fixed and the dimer length is altered by adjusting the As coordinates, shrinking and compressing the Mn-As tetrahedra in the out-of-plane direction. We find two magnetic phases as the As-As dimer length varies: a low-moment phase of 0.01 $\mu_{B}$ ("dimer phase"), and a higher-moment phase with magnitudes reaching $\sim$ 0.12 $\mu_{B}$ for an As-As dimer length of $\sim$ 5.4 \AA\ ("broken dimer phase"), separated by a QCP. This is in agreement with the alloy calculations in Fig.\ref{rbba_only}, demonstrating that this dimer length dictates the As-p magnetism.

\begin{figure}
 \centering
 \includegraphics[width=\linewidth]{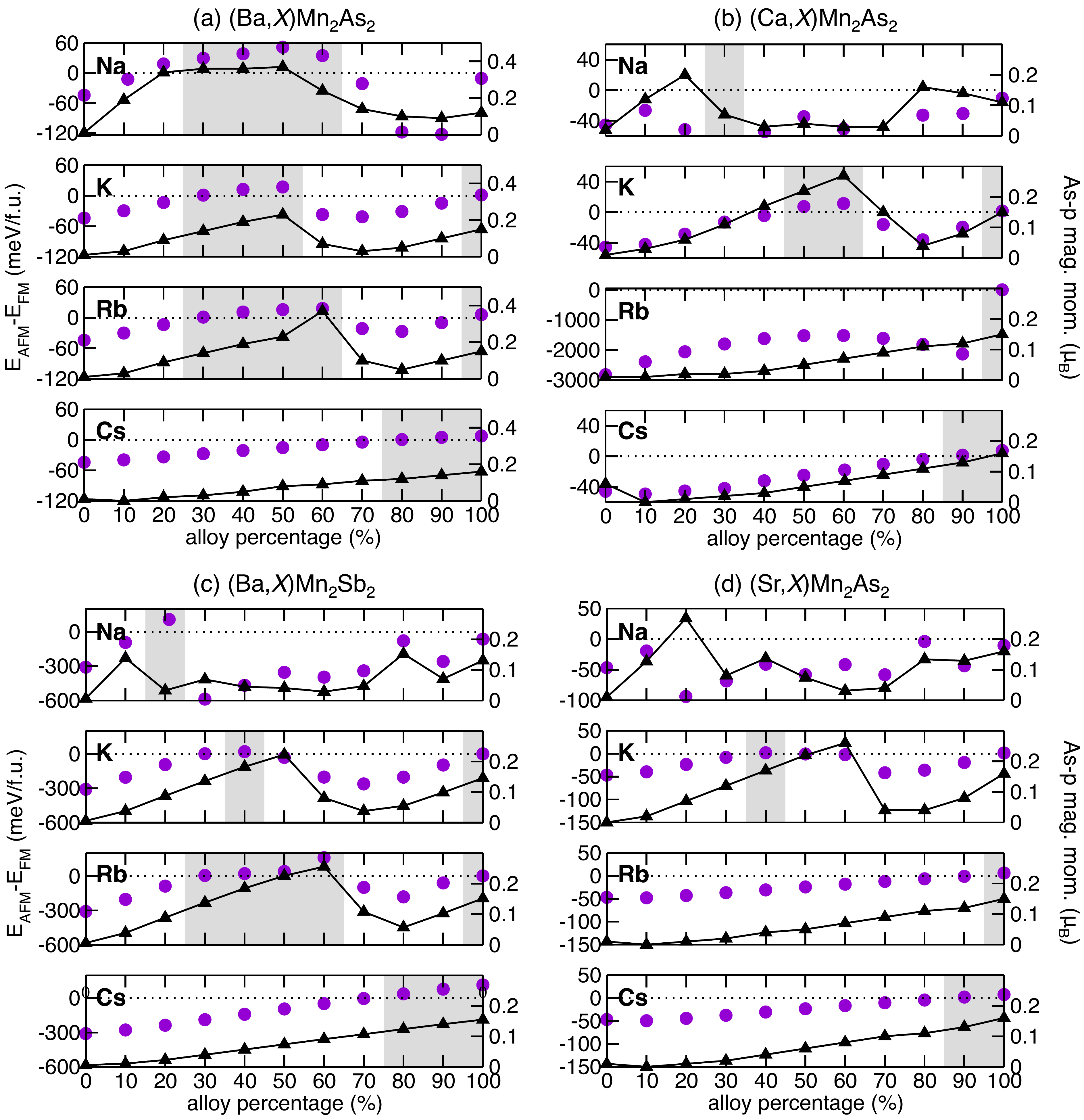}
 \caption{The circles (left axis) show the calculated energy difference between G-type AFM and FM orders for (a) (Ba,\textit{X})Mn$_{2}$As$_{2}$, (b) (Ca,\textit{X})Mn$_{2}$As$_{2}$, (c) (Ba,\textit{X})Mn$_{2}$Sb$_{2}$ and (d) (Sr,\textit{X})Mn$_{2}$As$_{2}$ where \textit{X}=Na, K, Rb, Cs. The triangles show the corresponding As or Sb magnetic moments (right axis). The shaded regions correspond to those alloy percentages with a FM ground state.}
 \label{combined}
\end{figure}

Following this intuition, we expand our search for HMFM in the I4/mmm structure by calculating a selection of group I and group II element combinations on the Ba site -- namely (Ba,\textit{X})Mn$_{2}$As$_{2}$, (Sr,\textit{X})Mn$_{2}$As$_{2}$, (Ca,\textit{X})Mn$_{2}$As$_{2}$ and (Ba,\textit{X})Mn$_{2}$Sb$_{2}$ alloys with \textit{X}=Na, K, Rb, Cs. The fully-relaxed structural parameters are given in SI (III), with the results for the various alloy combinations given in Fig.\ref{combined}. The shaded region in each shows the alloy percentages with a FM ground state. We find that the ground-state magnetic order for the stoichiometric compounds is always checkerboard AFM -- therefore all magnetic calculations compare this AFM to FM order. For each family we find some alloy concentration having a FM ground state -- with the largest regions in (a) (Ba,\textit{X})Mn$_{2}$As$_{2}$ and (c) (Ba,\textit{X})Mn$_{2}$Sb$_{2}$. As found in Ba$_{1-x}$Rb$_{x}$Mn$_{2}$As$_{2}$, the FM ground state correlates to the onset of substantial As-p or Sb-p moments. We note that the appearance and onset of FM changes for different parent compounds because of the different structural parameters (in particular the As-As bond lengths) of the parent compounds.

We address the feasibility for synthesis by considering the energetic stability of the stoichiometric compounds (0\% and 100\%) along with 50\% alloys in the desired I4/mmm structure and the P$\bar{3}$m1 structure -- the known experimental ground state structure of CaMn$_{2}$As$_{2}$. We find an I4/mmm ground state for BaMn$_{2}$As$_{2}$ and BaMn$_{2}$Sb$_{2}$. Experiments to date on Ba$_{0.4}$Rb$_{0.6}$Mn$_{2}$As$_{2}$\cite{Pandey/Johnston:2015} and Ba$_{0.6}$K$_{0.4}$Mn$_{2}$As$_{2}$\cite{Pandey_et_al:2013} suggest that the Ba compounds are robust in the I4/mmm structure up to large alloy concentrations. However CaMn$_{2}$As$_{2}$ and SrMn$_{2}$As$_{2}$ -- with smaller $A$-site cations -- adopt the P$\bar{3}$m1 structure. Alloying in the larger group I ions (50\% Rb and 50\% Cs), however, stabilizes the Ca- and Sr-based alloys in the I4/mmm structure. Though these materials could also potentially form in other stoichiometries, we do not consider the analysis of such binary and ternary compounds in this work since experimentalists have already successfully synthesized high alloy concentrations  in the $I4/mmm$ structure for two members of these series -- Ba$_{0.4}$Rb$_{0.6}$Mn$_{2}$As$_{2}$ and Ba$_{0.6}$K$_{0.6}$Mn$_{2}$As$_{2}$.

In summary, our first-principles calculations show the proximity of BaMn$_{2}$As$_{2}$ to a half-metallic electronic structure (Fig. \ref{ba_stoich}), and, using the virtual crystal approximation, we show that alloying with alkali metals provides the required hole doping to achieve half-metallicity with minimal spin contamination.

Experiments on Ba$_{0.6}$K$_{0.4}$Mn$_{2}$As$_{2}$ and Ba$_{0.4}$Rb$_{0.6}$Mn$_{2}$As$_{2}$ suggest them to be HMFM with $T_{C}$s of 100 K\cite{Pandey_et_al:2013} and 103 K\cite{Pandey/Johnston:2015}, respectively. Our results suggest that $T_{C}$ could potentially be increased by alloying into regions with a greater $\Delta$E. We show the ordered FM moment due to the As-p states to be 0.19 $\mu_{B}$ for Ba$_{0.6}$K$_{0.4}$Mn$_{2}$As$_{2}$ and  0.37 $\mu_{B}$ for Ba$_{0.4}$Rb$_{0.6}$Mn$_{2}$As$_{2}$, which slightly underestimate the respective experimental values of 0.45$\mu_{B}$ and 0.64$\mu_{B}$. 

The origin of ferromagnetism in these materials is a vital question to understanding appearance of half-metallicity. First-principles previously attributed this in-plane magnetization to a canting of the Mn-d moments\cite{Glasbrenner/Mazin:2014}. However, as evident in Fig.\ref{rbba_only} and Fig.\ref{combined}, the FM moments due to the As-p states are best correlated with the bond length of the intralayer As-As dimer, analogous to the Ge-Ge dimer used to explain the onset of FM in SrCo$_{2}$(Ge$_{1-x}$P$_{x}$)$_{2}$\cite{Jia_et_al:2011}. As the dimer length increases over $\sim$ 4 \AA, the As-p moments increase above 0.2 $\mu_{B}$. Alloys exhibiting robust ferromagnetism (As-p moments above $\sim$ 0.2 $\mu_{B}$) undergo a sharp decrease in both As-As dimer length and As-p moment with increased alkali concentration. This type of lattice collapse in the ThCr$_{2}$Si$_{2}$ structure was first explained over three decades ago by Hoffmann and Zheng\cite{Hoffmann/Zheng:1985}. The pnictogen-pnictogen intralayer bonding reduces the spacing between the transition-metal-pnictogen layers resulting in a so-called `collapsed tetragonal'  (ct) phase. The breaking of this dimer, and the intralayer bonding it induces, results in an `uncollapsed tetragonal' (uct) phase\cite{Hoffmann/Zheng:1985, Just/Paufler:1996}. This transition between the ct and uct phases can be driven experimentally by external pressure and chemically-induced pressure to a variety of `122' materials\cite{Huhnt_et_al:1997, Chefki_et_al:1998, Canfield_et_al:2009, Friedemann_et_al:2010}. In our case, the decrease in both As-As bond length and As-p moment can be understood as a transition between the the uct and ct phases, leading to a ferromagnetic quantum critical point (QCP). This uct-ct transition also explains the deviations from Vegard's law\cite{Vegard:1921}, which are seen in those alloys that exhibit the ferromagnetic QCP (see SI (III)). Our analysis is consistent with recent experiments in Ba$_{0.6}$K$_{0.4}$Mn$_{2}$As$_{2}$ which use x-ray magnetic circular dichroism to show that FM arises from the doped holes on As\cite{Ueland_et_al:2015}.

The appearance of magnetic moments on As-p states can be explained by charge redistribution that occurs in going from the uct to the ct phases. Following intuition from the Zintl concept\cite{Hoffmann/Zheng:1985, Jia_et_al:2011}, in the stoichiometric ct phase, the As-As dimer has a formal electron count of 4-; however in the uct phase, the As-As dimer is no longer bonded leaving each As with a formal electron count of 3-. The resulting charge redistribution in the Mn$_{2}$As$_{2}$ layer induces the observed ferromagnetism. In our case, this simple consideration is complicated by the hole-doping from the alkali-metal alloying; however, we isolate this influence in Fig.\ref{qcp}, demonstrating this mechanism independent of hole-doping.

Achieving half-metallicity in materials based on the BaMn$_{2}$As$_{2}$ therefore has two requirements: hole doping is needed to shift the Fermi level and cause a half-metal; and stabilizing the uncollapsed tetragonal phase of the I4/mmm structure type is needed for ferromagnetic ordering. Considering these requirements and our alloying calculations, the most promising half-metals are in the (Ba,\textit{X})Mn$_{2}$As$_{2}$ and (Ba,\textit{X})Mn$_{2}$Sb$_{2}$ alloy series, both of which have the I4/mmm ground state, undergo the ct-uct phase transition and exhibit ferromagnetic itinerant carriers. Two of these alloys -- (Ba,K)Mn$_{2}$As$_{2}$ and (Ba,Rb)Mn$_{2}$As$_{2}$ -- have already been identified to exhibit half-metallicity. However, with our suggested mechanism, their performance can be optimized by increasing the alkali concentration, increasing the magnitude of the As-p moments and T$_{C}$.

This new class of systems exhibiting HMFM offers several advantages over existing (Ga,Mn)As systems; importantly the solubility issues affecting single-crystal Mn-substituted GaAs are avoided by beginning with bulk BaMn$_{2}$As$_{2}$. Monovalent cation substitution on the Ba site has been achieved to remarkable success in BaFe$_{2}$As$_{2}$\cite{Rotter_et_al:2008}, and more recently in Rb- and K-substituted BaMn$_{2}$As$_{2}$\cite{Pandey_et_al:2013, Pandey/Johnston:2015}. Moreover, the half-metallicity reported in (Ba$_{1-x}$K$_{x}$)(Zn$_{1-y}$Mn$_{y}$)As$_{2}$, requires both hole doping by (Ba$^{2+}$,K$^{1+}$) replacements and spin doping with (Zn$^{2+}$,Mn$^{1+}$) replacements\cite{Zhao_et_al:2013}. Our new class requires only hole doping via monovalent alloying where ionic size affects induce magnetism. Furthermore, with reported $T_{C}$s over 100 K, these materials offer up new routes to achieving room temperature HMFM.

This work was supported by the Director, Office of Science, Office of
Basic Energy Sciences, Materials Sciences and Engineering Division, of
the U.S. Department of Energy under Contract No. DE-AC02-05-CH11231.
Computational resources provided in part by the Molecular Foundry was
supported by the Office of Science, Office of Basic Energy Sciences,
of the U.S. Department of Energy, also under Contract No.
DE-AC02-05-CH11231. SG acknowledges financial support by the Swiss
National Science Foundation Early Postdoctoral Mobility Program.

\bibliography{sinead}

\end{document}